\documentclass[superscriptaddress,showpacs,aps,twocolumn,floatfix,prl,longbibliography]{revtex4-1}
\usepackage{bm,amsmath,amssymb}
\usepackage{bbm}
\usepackage{amssymb}
\usepackage{amsmath}
\usepackage{mathrsfs}
\usepackage{graphicx,bm}
\usepackage{subcaption}
\usepackage{verbatim}
\usepackage[dvipsnames]{xcolor}
\usepackage{soul}
\usepackage{ bbold }
\usepackage{xspace}
\usepackage{enumitem}

\usepackage[dvipsnames]{xcolor}
\usepackage[normalem]{ulem}

\usepackage{pgfmath}
\usepackage{pgfplots}
 
\pgfplotsset{compat = newest}

\usepackage[colorlinks]{hyperref}

\AtBeginDocument{%
 \hypersetup{ 
 linkcolor=teal,
 filecolor=magenta, 
 urlcolor=teal,
 citecolor=Maroon,
 colorlinks=true,
 }
}

\usepackage[T1]{fontenc}
\usepackage[utf8]{inputenc}

\setlist[itemize]{itemsep=0pt, topsep=3pt}

\newcommand{\id}{\mathbbm{1}}
\def\a{\alpha}
\def\b{\beta}

\newcommand{\p}	{\partial}

\newcommand{\Z}	{\mathbbm{Z}}

\newcommand{\cF}{\mathcal{F}}

\newcommand{\cH}{\mathcal{H}}
\newcommand{\cI}{\mathcal{I}}

\newcommand{\cO}{\mathcal{O}}

\DeclareMathOperator{\tr}{tr}

\newcommand{\bra}[1]	{\langle{#1}\vert}
\newcommand{\ket}[1]	{\vert{#1}\rangle}

\newcommand{\ketbra}[2]	{\ket{#1}\bra{#2}}
\newcommand{\corr}[1]   {\left\langle{#1}\right\rangle}

\newcommand{\bL}    {\bar{L}}
\newcommand{\bh}    {\bar{h}}

\newcommand{\btau}  {\bar{\tau}}

\newcommand{\tw}    {\tilde{w}}
\newcommand{\tq}    {\tilde{q}}
\newcommand{\ttau}  {\tilde{\tau}}

\newcommand{\cc}        {\mathsf{c}} %central charge
\newcommand{\gf}        {\mathsf{g}} %g-factor

\newcommand{\bbra}[1]	{\langle\!\langle{#1}\lVert}% negative space is produce by \!
\newcommand{\bket}[1]	{\lVert{#1}\rangle\!\rangle}
\newcommand{\iket}[1]	{\lvert{#1}\rangle\!\rangle}

\newcommand{\fus}       {\mathsf{N}}
\newcommand{\fusb}       {\mathsf{M}}
\newcommand{\fuse}      {\star}
\newcommand{\qdim}      {\mathsf{d}}

\renewcommand{\th}      {\text{th}}

\newcommand{\iu}{\mathsf{i}}%imaginary unit

\newcommand{\duality}   {\dutchcalD}

\newcommand{\ab}        { {\alpha\beta} }

\newcommand{\wi}        {\mathsf{W}} % width of annulus

\newcommand{\X}         {\mathsf{X}}

\newcommand{\Ssym}    {S^S}
\newcommand{\Sint}    {S^I}
\DeclareMathAlphabet{\mathdutchcal}{U}{dutchcal}{m}{n}
\SetMathAlphabet{\mathdutchcal}{bold}{U}{dutchcal}{b}{n}
\DeclareMathAlphabet{\mathdutchbcal}{U}{dutchcal}{b}{n}

\newcommand{\dutchcalD}{\text{\usefont{U}{dutchcal}{m}{n}D}} 

\newcommand\quotes[1]  {\lq\lq{#1}\rq\rq}
\renewcommand\emph[1]  {\textbf{#1}}

\newcommand\figref[1]	{figure~\ref{#1}\xspace}

\newcommand{\lref}[1]{\hyperref[#1]{(\ref*{#1})}}
\makeatletter
\newcommand{\itemlabel}[1]{%
  \phantomsection
  \def\@currentlabel{#1}%  <-- what \ref{...} should return
}
\makeatother

\begin{document}

\title{Entanglement Through Topological Defects: Reconciling Theory with Numerics}
% options for title: put in a finding, use positive words, two part titles (part one : part two)
% avoid: noun stacking, delete trivial stuff like "a study of..."

\author{Christian Northe}
\author{Paolo Rossi}
\thanks{northe@fzu.cz}
\thanks{rossip@fzu.cz}
\affiliation{CEICO, Institute of Physics of the Czech Academy of Sciences,\\
Na Slovance 2, 182 00 Prague 8, Czech Republic}

% \date{April 2022}

\begin{abstract}
% 1) state the problem
% 2) why is this paper written
% 3) what is done, methods
% 4) findings (the main findings are)
% 5) conclusions, significance, who benefits, what do results mean in broader context?
% 6) make sure sentences direct a reader to one of the five points above. 
% 7)prefer short sentences. if they need to be read twice, change them! "spoon feed the readers" by making things explicit
%8) is gap indicated (although, however, despite... )
% 9) are the main keywords in the first two sentences? 
Present theoretical predictions for the entanglement entropy through topological defects are violated by numerical simulations. In order to resolve this, we introduce a paradigm shift in the preparation of reduced density matrices in the presence of topological defects, and emphasize the role of defect networks with which they can be dressed. We consider the cases of grouplike and duality defects in detail for the Ising model, and find agreement with all numerically found entanglement entropies. Since our construction functions at the level of reduced density matrices, it accounts for topological defects beyond the entanglement entropy to other entanglement measures.
\end{abstract}

\maketitle
%%%%%%%%%%%%%%%%%%%%%%%%%%%%%%%
\textit{1. Introduction}.---  Defects and interfaces appear ubiquitously in nature, from impurities in materials \cite{AffleckGFactor} over quantum dots \cite{sierra2014strongly} and tensor networks \cite{Hauru:2015abi} all the way to D-branes in string theory \cite{Polchinski:1998rr}. The last decade has, moreover, witnessed a reformulation of our understanding of symmetries to what is now known as generalized symmetry \cite{Shao:2023gho} -- a development driven mainly by topological defects. These are completely energy-transmissive and can therefore be deformed at will, until fields or boundaries are encountered, where topological defects act in various interesting ways. In addition to implementing conventional group actions, topological defects can implement, for instance, order-disorder dualities \cite{Frohlich_2007}, most famously Kramers-Wannier duality in the Ising model \cite{Frohlich:2004ef}. 

For these reasons, it is important to understand how topological defects affect quantum correlations. A natural candidate for their quantification is the entanglement entropy (EE). In zero-temperature ground states of 1+1D quantum-critical systems described by a conformal field theory (CFT), EE depends universally on the central charge~$\cc$ \cite{Calabrese_2009}. Since $\cc$ labels the universality classes of CFTs \cite{Zamolodchikov:1986gt}, EE becomes a powerful tool for classifying such systems. Boundaries contribute at subleading order to the EE \cite{Calabrese_2009} through the Affleck-Ludwig boundary entropy $\gf$ \cite{AffleckGFactor}, which parallels $\cc$ in that it classifies universality classes of conformal boundary conditions \cite{Friedan:2003yc}. 

The effect of topological defects, in contrast, on entanglement is more nuanced. They appear for instance naturally as a tool for symmetry resolution of entanglement spectra \cite{Belin_2013, Goldstein:2017bua, Zhao_2021, Weisenberger_2021, Zhao:2022wnp,  northe2023entanglement, Kusuki:2023bsp, Choi:2024wfm, Kusuki:2024gss, Heymann_2025, Northe:2025qcv, benini2025entanglementasymmetryhighernoninvertible, Bhattacharyya:2025tmg, Saura-Bastida:2024yye, Das:2024qdx, magan2021proof, capizzi2022symmetry, Capizzi:2022igy}, where the defects are imposed manually. Situations where the defect forms part of the physical system, crossing in particular through one entangling edge, were studied early on \cite{Sakai_2008, Brehm:2015lja, Brehm_2016,Brehm:2017kuz, Gutperle_2016} in high-energy theory. Their proposed framework and results are widely accepted beyond high energy physics in quantum information and condensed matter \cite{Eisler:2012xry, Wen:2017smb, Afxonidis:2024gne, Gutperle:2017enx, gutperle2024noteentanglemententropytopological, capizzi2023domain, Capizzi:2023mly, Capizzi:2023vsz}. Said framework \cite{Sakai_2008, Brehm_2016, Gutperle_2016} predicts a subleading contribution to the EE, which, however, has recently been seen to be contradicted in \textit{ab initio} numerical simulations of the Ising model \cite{Roy_2022, Rogerson:2022yim}, the 3-state Potts model \cite{Sinha:2023hum} and the Luttinger liquid \cite{Roy:2023wer}. 

Subleading contributions encode the intricate interplay between boundaries, defects, but also field excitations \cite{Ibáñez_Berganza_2012} and the entanglement spectrum \cite{Li:2008kda, Qi:2012ngv}. The latter is the eigenspectrum of the reduced density matrix (RDM) and captures therefore much more information than the EE. It is thus important to have a formalism capable of predicting the effect of topological defects on quantum correlations correctly. The purpose of this letter is to fill this gap.

It is useful to begin with a sketch of the prevailing theoretical framework \cite{Sakai_2008}.
Consider an interface $L$ separating the right-half from the left-half plane, and denominate the positive real line as entangling interval $A$. A reduced density matrix $\rho_A^{old}$ is prepared by identifying two concentric circles centered at the entangling edge $\p A$, one with radius $\epsilon\ll1$, and one with radius $1/\epsilon$,
\begin{align}\label{eq:RDMold}
\vcenter{\hbox{\includegraphics[height=2.4cm]{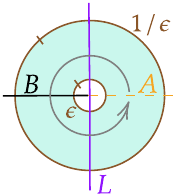}}} \quad \Rightarrow\quad \rho^{old}_A \propto \vcenter{\hbox{\includegraphics[height=2.3cm]{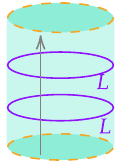}}}
\end{align}
In this way the interval $A$ is compactified onto a circle, and the RDM is a defect-dressed cylinder in which entanglement time runs from bottom to top. Because the identification splits the defect in two, it acts twice on the subregion Hilbert space $\cH_A$. Stacking $n$ such RDMs and tracing over $\cH_A$ leads to a torus. Although, this framework produces the correct leading order in the EE, two problems are apparent:
\begin{itemize}[itemsep=2pt, parsep=0pt, partopsep=0pt]
    \item[(P1)] \itemlabel{P1} \label{P1} The topology of the interval and the defect network are no longer faithful to the original configuration. 
    \item[(P2)] \itemlabel{P2} \label{P2} The entanglement spectrum $\cH_A$ has left- and right-movers in this framework, standing in conflict with more recent developments, which establish a chiral entanglement spectrum for a single interval \cite{Lauchli:2013jga, Cardy_2016, Cho:2016kcc, Roy_2020, Hu:2020suv}.
\end{itemize}
The RDM \eqref{eq:RDMold} predicts subleading terms in the EE \cite{Brehm_2016, Gutperle_2016}, which depend explicitly on two defects $L$ instead of one, due to \lref{P1}, apply to a compact subregion, instead of a line segment, again due to \lref{P1}, and draw on its non-chiral entanglement spectrum, due to \lref{P2}. Precisely these predictions are violated in the simulations \cite{Roy_2022, Rogerson:2022yim, Sinha:2023hum, Roy:2023wer}. A further element of contention is that the simulated subleading terms depend generically on the subregion size, which is not mirrored by the theory of \cite{Sakai_2008}. 

In this letter, we construct RDMs for an interval $A$, which do not suffer from these problems and reproduce the simulations in the Ising model \cite{Roy_2022}. Besides having $L$ pierce the entangling edge $\p A$, we also discuss the situations where $L$ crosses inside and outside of $A$. Our guiding principle is to consider proper choices of defect network that complete the defect piercing the constant time slice. Ultimately, this leads us to introduce twist fields and study their quantum correlations. We work directly at the level of RDMs, so that our formalism's utility extends beyond the EE to the investigation of other entanglement measures. 

After briefly recapitulating boundaries and defects, we construct our RDMs and match their R\'enyi entropies to numerically obtained EEs in the Ising model \cite{Roy_2022}.

%%%%%%%%%%%%%%%%%%%%%%%%%%%%%%%%%%%
\textit{2. Boundaries and defects}.---
We shall work with CFTs with diagonal modular invariants, e.g. the A-series minimal models, of which the Ising CFT is the simplest one. Their conformal boundary conditions $\a$, i.e. those preserving conformal symmetry along their locus, are labeled by the set of conformal families $\a\in\cI$ and are described by Cardy states $\bket{\alpha}$. In the Ising CFT, there are only three families $\cI_I=\{0,1/2,1/16\}$. Topological defects $D_a$, called Verlinde lines, are labeled by the same set, $a\in\cI$, and commute with all generators of conformal transformations. Thus they can be deformed at will, so long as no field insertion is crossed.

Verlinde lines satisfy a fusion algebra $D_aD_b=\sum_c\fus_{ab}^cD_c$, where the $\fus_{ab}^c\in\{0,1\}$ are the fusion coefficients. We are interested in two types of defects \cite{Frohlich:2004ef, Frohlich_2007}: 1) Grouplike defects $D_g$ implement a group action $g\in G$ and have an inverse $D_{g^{-1}}D_g=D_gD_{g^{-1}}=\id$. In the Ising CFT these are $D_0=\id$ and $D_{\varepsilon}$. Because $D_\varepsilon$ realizes a domain wall between regions of positive and negative spins, the group in question is the spin-flip $\Z_2$ symmetry. 2) Duality defects $\duality$ implement order-disorder dualities. They fuse with their orientation inverse into grouplike defects, $\duality^\dagger\duality=\sum_{g\in G}D_g$. In the Ising CFT, this role is taken by $D_{\sigma}$, which satisfies $D_\sigma D_\sigma=D_1+D_\varepsilon$, and implements Kramers-Wannier duality \cite{Frohlich:2004ef}.

Defects $D_a$ can terminate on twist fields in the bulk \cite{Petkova:2000ip, Northe:2024tnm}. The standard disorder field $\mu$ with dimensions $h=\bh=1/16$ in the Ising model is the lowest-energy field terminating $D_\varepsilon$. The ground state for twist fields terminating $D_\sigma$ is shared by $\nu$ with $h=1/16$, $\bh=0$ and $\bar{\nu}$ with $h=0$, $\bh=1/16$. Defects can furthermore host fields. For instance any $D_a$ supports an identity field $\id_a$ with $h=\bh=0$~\cite{Petkova:2000ip, Northe:2024tnm}.

Verlinde lines act on Cardy states by fusion $D_a\bket{\alpha}=\sum_b\fus_{a\a}^{\b}\bket{\b}=:\bket{a\fuse\a}$ with fusion coefficients $\fus$. Hence defects can end on a boundary, and, for their junction to be topological, it suffices to demand $\beta\in a\fuse\a$,
\begin{align}\label{eq:fusing_line_bound}
\vcenter{\hbox{\includegraphics[height=1.3cm]{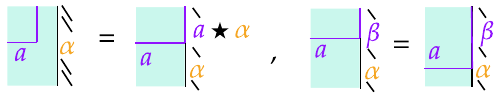}}}
%\raisebox{-.45\height}{\includegraphics[scale=.2]{Pics/BdyDefMoves.png}}
\end{align}
Note that in the second picture the defect can only be unfused if $\beta=a\fuse\a$. Lastly, the Affleck-Ludwig boundary entropy $\gf_{\a}=\langle0\bket{\a}$ \cite{AffleckGFactor} changes upon application of a Verlinde line to $\gf_{a\fuse\a}=\bra{0}D_a\bket{\a}=\qdim_a\gf_\a$, where $\qdim_a$ is the quantum dimension.

\begin{figure}
%\captionsetup{justification=raggedright, singlelinecheck=false}
\subfloat[\label{fig:RDM_state}]{\includegraphics[height=2.5cm]{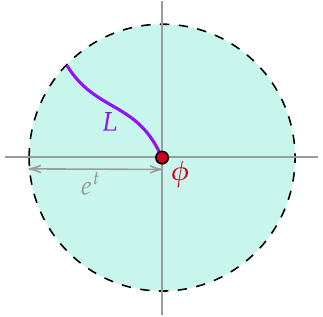}} \hspace{0.5cm}
\subfloat[\label{fig:RDM_OT}]{\includegraphics[height=2.7cm]{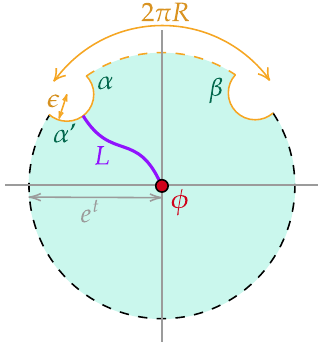}}\hspace{0.5cm}
\subfloat[\label{fig:RDM_spherestrip}]{\includegraphics[height=2.3cm]{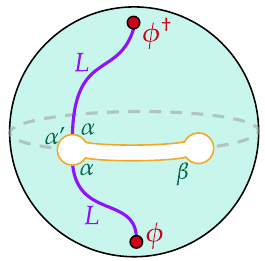}}
\caption{\small (a) Preparation of the twisted state $\ket{\phi}(t)$ in presence of the topological line $L$ (radial quantization). (b) The state is mapped to ${\cal H}_{\ab}^A\otimes {\cal H}_{\b\a'}^B$ by excising disks and imposing boundary conditions. Case ($ii$) is picked as an example. The length $2\pi R$ of the region $A$ is measured by a dimensionless parameter $R\in(0,1)$, so that $R=0,1$ correspond to vanishing $A$ or $B$, respectively. (c) RDM $\rho^{\phi}_{\ab}$ obtained from (b) after tracing over $B$.
%\change{The radial quantization (sphere) frame is mapped to a strip of sides ratio $\approx(1:\frac{|\log \epsilon|}{\pi})$ after a conformal transformation, setting the distance between the punctures to $R$.}{Put it somewhere else}
}
\label{fig:RDMconstruction}
\end{figure}

%%%%%%%%%%%%%%%%%%%%%%%%%%%%%%%%%%%
\textit{3. RDMs in the presence of defects}.---
The physical situation we consider is that of a circular constant time slice pierced by a topological defect. To be unambiguous, this setup needs to be completed by assigning a proper defect network. In \eqref{eq:RDMold} this is achieved by closing the defect on itself through compactification -- at the price of problems \lref{P1} \& \lref{P2}. The route we shall pursue here instead, is to have the defect emanate from a twist field $\phi$ prepared in the distant past, see \figref{fig:RDM_state}. Two significant modifications compared to $\rho_A^{old}$ arise:
\begin{itemize}[itemsep=2pt, parsep=0pt, partopsep=0pt]
    \item[(M1)] \itemlabel{M1} \label{M1} The twist field $\phi$ is not part of the modular invariant describing the global spectrum $\cH$. We must therefore consider  $\cH$ together with all twisted sectors.
    \item[(M2)] \itemlabel{M2} \label{M2} The state $\phi$ is not the global vacuum of the theory, and we therefore study quantum correlations in an excited state.
\end{itemize}
It is the goal of this letter to demonstrate that \lref{M1} \& \lref{M2} are feature, not a bug, and in fact essential in reproducing the numerical observations in the Ising model \cite{Roy_2022}.
We now proceed to construct RDMs based on these modifications.

In the absence of additional structure, it does not matter where the topological defect pierces the constant time slice. However, when declaring a subregion $A$, its position relative to the line $L$ becomes relevant. Three cases are distinguished: $L$ crosses ($i$) within the interval $A$, ($ii$) at one of its boundary points, or ($iii$) outside of $A$,
\begin{center}
    \includegraphics[height=2.5cm]{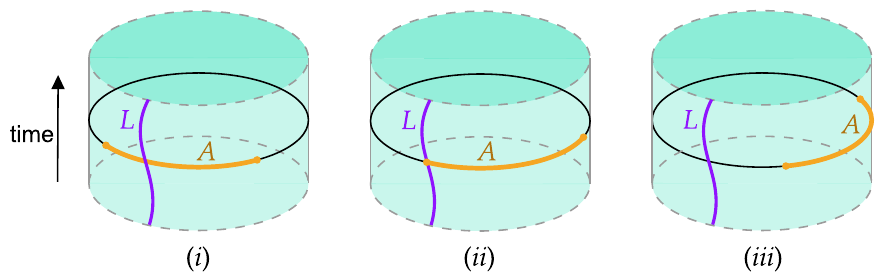} 
\end{center}
The EE associated with $(i)$ or $(iii)$ accounts for quantum correlations in an impurity-doped system $A$ or $B$, respectively. They are called \quotes{symmetric entropy} for $A$ or $B$. The EE of $(ii)$, on the other hand, measures quantum correlations across the defect, and is called \quotes{interface entropy.}

In order to factorize the Hilbert space $\cH\to\cH_{\ab}^{A}\otimes\cH_{\b\a}^{B}$, we excise a disk surrounding the entangling edges and impose boundary conditions $\a,\,\b$ thereon \cite{Ohmori_2015}. In case ($ii$), $L$ attaches to the boundary $\a$, say. As explained in \eqref{eq:fusing_line_bound}, this is possible only if the fusion rules allow it and a-priori requires the choice of another boundary condition $\a'\in L\star\a$, see \figref{fig:RDM_OT}. An RDM is constructed out of this state and its adjoint as in \figref{fig:RDM_spherestrip}. Finally, a conformal map allows us to depict RDMs as strips,
\begin{equation}\label{DefectRDMs}
\rho_{\alpha\beta}^{\phi,S} \propto \vcenter{\hbox{\includegraphics[height=1.8cm]{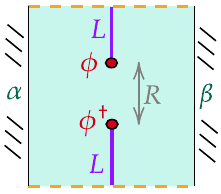}}}\,
\qquad
\rho_{\alpha\beta}^{\phi,I} \propto \vcenter{\hbox{\includegraphics[height=1.8cm]{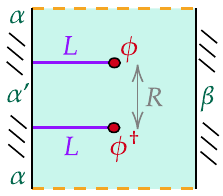}}}
\end{equation}
up to a normalization enforcing ${\rm Tr}\rho^\phi_{\ab}=1$. The strip's height is 1 and its width $\frac{1}{\pi}|\log \ell(R)/\epsilon|$, with $\ell(R)=2\sin (\pi R)$, is controlled by the interval size $R$ defined in \figref{fig:RDM_OT}. The punctures are a distance $R$ apart and entanglement time runs from bottom to top. $\rho^{\phi,S}_\ab$ belongs to the \quotes{symmetric} case $(i)$ and $\rho^{\phi,I}_\ab$ to the \quotes{interface} case $(ii)$. Replica manifolds for case $(iii)$ are equivalent to $\rho_\ab^{\phi,S}$ with $R\to 1-R$ and are not treated separately here. We stress that the two RDMs in \eqref{DefectRDMs} are not distinguished by the conformal weights $h_\phi$ and $\bh_\phi$ of the twisted primary, but \textit{only by the defect network}. This proves to be crucial, especially when dealing with entanglement in twisted states associated with non-invertible defect lines. All details and our conventions for the factorization procedure are found in \cite{Northe:2025qcv}.

Note that our framework manifestly avoids the problems \lref{P1} \& \lref{P2}. Indeed, the topology of the interval $A$ and the defect network in the original configuration stay intact. Furthermore, the RDMs \eqref{DefectRDMs} act on a chiral Hilbert space $\cH_\ab^A$, which provides the entanglement spectrum. The main message of our letter is thus that \textit{entanglement through defects is captured by the RDMs \eqref{DefectRDMs}} instead of the RDM \eqref{eq:RDMold}.

For an untwisted global primary state $\ket{\phi}$, the R\'enyi entropy is found to be \cite{Ibáñez_Berganza_2012, Ohmori_2015, Northe:2025qcv}
\begin{equation}\label{EE}
S_{n}(\rho_{\ab}^{\phi})
=
\frac{1+n}{n}\frac{\cc}{6}\log\left(\frac{\ell(R)}{\epsilon}\right)
+
\log[\gf_{\a}\gf_{\b}\,\cF_n^\phi(R)]+\cdots
\end{equation}
where $\cc$ is the central charge of the CFT and $\cF_n^\phi(R)$ results from $\phi^{\times 2n}$ correlators, e.g. for the vacuum field, $\cF_n^\id(R)=1$. While a rescaling of the cutoff $\epsilon$ can absorb the $\gf_\a\gf_b$ term, it can always be recovered by comparing entropies of distinct $n$. Dots stand for vanishing contributions in the limit $\epsilon\to 0$. 

We now carry on to reproduce \eqref{EE} from the RDMs \eqref{DefectRDMs} of \textit{twisted} primary states in the Ising CFT, first for grouplike defects, and thereafter for duality defects. In both cases we find agreement with the numeric results of \cite{Roy_2022}, thereby validating the modifications \lref{M1} \& \lref{M2}, and thus the RDMs \eqref{DefectRDMs}. 

%------------------------------
\textit{4. Matching EE for grouplike defects}.--- 
Numeric modeling of the symmetric and interface entropy results in the following ground state symmetric (S) and interface (I) EEs for the grouplike defect $D_\varepsilon$ \cite{Roy_2022} 
\begin{align}\label{NumEEgrouplike}
\Ssym_1(D_\varepsilon)
=
\Sint_1(D_\varepsilon)
=
S_1(\id)+\cO(\epsilon)
\end{align}
where $S_1(\id)$ is the EE in the global vacuum. We now demonstrate that this result is reproduced from \eqref{EE} when choosing the RDMs \eqref{DefectRDMs} for the disorder field $\phi=\mu$. In fact, we show that both, $\rho_\ab^{\mu,S}$ and $\rho_\ab^{\mu,I}$, yield the same R\'enyi entropy as $\rho_\ab^\id$, not just EE.

We depict only $n=2$ replica manifolds, although our arguments hold for any $n$. A replica geometry for $\rho_\ab^{\mu,I}$ can be deformed into one for $\rho_{\a\fuse\varepsilon,\b}^{\mu,S}$. Indeed,
\begin{equation}\label{replicaGrouplikeMove}
\vcenter{\hbox{\includegraphics[height=2.5cm]{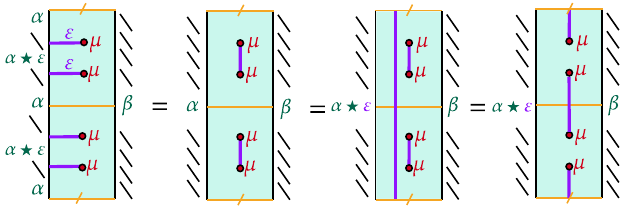}}}
%    \raisebox{-.45\height}{\includegraphics[scale=.178]{Pics/ReplicaGrouplike.png}}
\end{equation}
On the left, the only choice for $\a'$ is $\a\fuse\varepsilon$, because grouplike defects map elementary boundaries into themselves. Hence, the defect $D_\varepsilon$ can be trivially unfused from the boundary. In the first equality, this is done partially, and from the entire left boundary in the second equality, leaving $\a\fuse\varepsilon$. Using $D_\varepsilon^2=\id$, the defect line is moved onto the field insertions. 

At the price of dividing by the quantum dimension $\qdim_\sigma$, we nucleate a $D_\sigma$ loop in the second and fourth image of \eqref{replicaGrouplikeMove}. Moving it past the field insertions, $\mu$ turns to $\sigma$, shown exemplarily starting from the second image,
\begin{equation}\label{ReplicaDualityMove}
\vcenter{\hbox{\includegraphics[height=2.3cm]{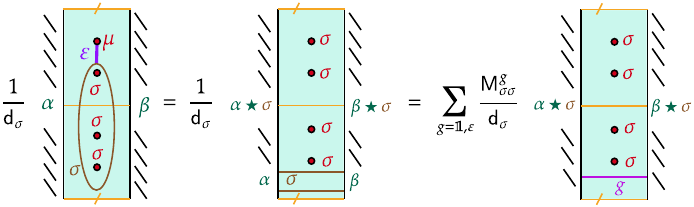}}}
%    \raisebox{-.45\height}{\includegraphics[scale=.165]{Pics/ReplicaDuality.png}}
\end{equation}
In the first equality, $D_\sigma$ is fused onto the boundaries. The structure constants $\fusb_{\sigma\sigma}^g$ encode the fusion of the two $D_\sigma$ lines into grouplike defects $D_g$ in the presence of boundaries \cite{Kojita:2016jwe}. Following conventions of \cite{Frohlich_2007}, $\fusb_{\sigma\sigma}^\id=1/\qdim_\sigma$. When repeating this move on the fourth image of \eqref{replicaGrouplikeMove}, the left boundary becomes $\a\fuse\varepsilon\fuse\sigma=\a\fuse\sigma$. Therefore the replica manifolds of $\rho_\ab^{\mu,S}$ and $\rho_\ab^{\mu,I}$ are both equal to the right-hand side of \eqref{ReplicaDualityMove}.

The $\epsilon\to0$ regime is dominated by the $g=\id$ summand, in which the boundaries move out to $\pm\infty$ leaving behind only 
$\fusb_{\sigma\sigma}^\id\gf_{\a\fuse\sigma}\gf_{\b\fuse\sigma}/\qdim_\sigma=\gf_{\a}\gf_{\b}$ times a $\sigma^{\times2n}$ correlator on the sphere. The moments are thus
\begin{equation}\label{EEmu}
    \tr_\ab[(\rho_\ab^{\mu,I})^n]
=
\tr_\ab[(\rho_\ab^{\mu,S})^n]
\simeq
%\tr_\ab[(\rho_\ab^{\sigma})^n]
%=
\tr_\ab[(\rho_\ab^{\id})^n]\cF_n^\sigma(R)\,.
\end{equation}
Due to $\cF_n^\sigma(R)=1$ \cite{Ibáñez_Berganza_2012}, the observation \eqref{NumEEgrouplike} is theoretically confirmed and extended to any $n$. All details are found in the supplemental material.

Notice that our analysis is independent of a particular choice of $\a,\b$. Furthermore, the defect deformations we employ are fairly generic and can be repeated in other models for any grouplike defect that lies in the stabilizer of a duality defect. Finally, because grouplike defects can be unfused from the boundary, interface entropies on the interval $A$ for grouplike defects can always be interpreted as symmetric entropies for $B$.

%%%%%%%%%%%%%%%%%%%%%%%%%
\textit{5. Matching EE for duality defects}.--- Simulations for the lattice ground state EE in the presence of the duality defect $D_\sigma$ find \cite{Roy_2022}
\begin{equation}\label{eq:numerics_sigma}
\begin{aligned}
    S_1^S(D_\sigma) &= S_1(\id) +f(R), \\
    S^I_1(D_\sigma) &= S_1(\id) +f(1-R) + \delta_{R,1}\frac{\log2}{2}
\end{aligned}
\end{equation}
where $f(R)$ is a known function \cite{Klich:2015ina}. Its specific form is not necessary for us, only its asymptotic values $f(R)\approx \frac{\log2}{2}$ for $R\to1$ and $f(R)\approx0$ for $R\to0$. Our framework breaks down at $R=0$ or $1$, since the usual conformal transformations to the replica manifold become singular. So we are not able to test the contribution $\frac{\delta_{R,1}}{2}\log2$ which would however stem from completing the EE to a regular entropy. It is important to recall that $S_1(\id)$ is invariant under $R\to 1-R$, so that $S_1^S(D_\sigma)$ and $S_1^I(D_\sigma)$ are exchanged under this swap (away from $R=1$).

When looking at the EE in the $\sigma$-twisted sector, we come across the subtlety that there are two choices of lowest lying states $\nu,\bar\nu$.
%, respectively of conformal dimensions $(1/16,0)$ and $(0,1/16)$. 
A priori one can consider any mixture of these, but we settle for a system prepared in the pure state $\nu$ \footnote{Rényi entropies are not expected to be a good measure of entanglement in mixed states \cite{Calabrese_2009}, which is why we do not consider e.g. the maximally mixed state $\frac{1}{\sqrt{2}}(\ket{\nu}+\ket{\bar\nu})$.}.

The non-invertible nature of $D_\sigma$ makes it difficult to find a network manipulation relating replica manifolds of $\rho_\ab^{\nu,S}$ and $\rho_\ab^{\nu,I}$ similar to \eqref{ReplicaDualityMove}. Furthermore, no bosonized representation for the field $\nu$ is known to us with which we could parallel the analysis in \cite{Ibáñez_Berganza_2012}. Nevertheless, we provide strong evidence in favor of the RDMs \eqref{DefectRDMs}:
%\change{slicker}{more direct?}\cn{actually, i would say \quotes{indirect}}: what we shall 
we show that, for a specific choice of intermediary boundary condition $\a'$ in \eqref{DefectRDMs}, $S_n(\rho_\ab^{\nu,S})$ and $S_n(\rho_\ab^{\nu,I})$ are related by $R\to1-R$, and furthermore that they reproduce the asymptotic values of $f(R)$ for small, $R\ll1$, and large intervals, $1-R\ll1$. These three non-trivial matches establish high confidence that the RDMs \eqref{DefectRDMs} indeed describe the $D_\sigma$-dressed subsystem.

Consider the symmetric RDM depicted in \eqref{DefectRDMs} for $\phi=\nu$ and $L=D_\sigma$. In the associated replica manifold, pairs of $\nu$ punctures at a distance $1-R$ are connected by $D_\sigma$,
\begin{equation} \label{eq:RDMsigma_symmetric}
\vcenter{\hbox{\includegraphics[height=1.6cm]{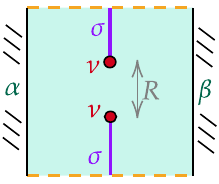}}}\quad \cong \quad \vcenter{\hbox{\includegraphics[height=1.6cm]{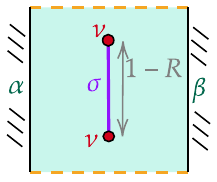}}} 
\end{equation}
For the interface entropy, we face the important subtlety that one can potentially choose different intermediate boundary conditions $\a'\in \sigma\star\a$ in the RDM $\rho^{\nu,I}_{\ab}$. We claim that the most natural choice, which reproduces the numerical results, is the proper fusion $\a'=\sigma\star\a$. Notice that if $\a=\sigma$, then $\a'=\id+\varepsilon$ is a non-elementary boundary condition, which poses no problem, however. With this choice, the two $D_\sigma$ lines can be unfused from the boundary similarly to the first move in \eqref{replicaGrouplikeMove}. The resulting RDM looks like the right-hand side of \eqref{eq:RDMsigma_symmetric} but with $1-R$ replaced by $R$. This shows $S_n(\rho_\ab^{\nu,S},R)=S_n(\rho_\ab^{\nu,I},1-R)$, as claimed. We remark that this type of argument does not apply in the conventional twist field picture of entanglement \cite{Calabrese_2009}, since detaching the defect from the entangling edge must be justified by alternate means.

To complete our argument, we now match the asymptotic values of the EE in the symmetric case. As $R\to 0$ the two $\nu$ punctures approach each other head first. We can use the OPE $\nu(z)\nu(w)= \frac{\id_\sigma}{(z-w)^{1/8}} + \cdots$, where $|z-w|=R$ and the dots vanish as $z\to w$. Because the defect tails connect into a continuous line, $\id_\sigma$ is the identity field on $D_\sigma$. To leading order in $1/R$ \eqref{eq:RDMsigma_symmetric} simplifies to
\begin{equation}
    \rho^{\sigma,S}_{\ab}\stackrel{R\ll1}{\approx} \frac{R^{-1/8}}{\cal N} 
    \vcenter{\hbox{\includegraphics[height=1.6cm]{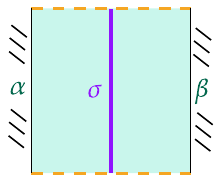}}}
     = \frac{R^{-1/8}}{\cal N'} \rho^{\id}_{\a , \sigma\star\b}
\end{equation}
where in the last step we fused the line to the right boundary and included normalizations $\cal N, N'$. The overall prefactor $R^{-1/8}$ cancels out due to the normalization and does not contribute to  the Rényi entropy. This shows that $S_n(\rho^{\sigma,S}_{\ab})\stackrel{R\ll1}{\approx} S_n(\rho^{\id}_{\a,\sigma\star\b})$. Using \eqref{EE} and the relation $\gf_{a\star \b}=d_a \gf_{\b}$, together with $d_\sigma = \sqrt{2}$ yields the expected result $S_n(\rho^{\sigma,S}_{\ab})\stackrel{R\ll1}{\approx} S_n(\rho^{\id}_{\ab}) + \frac{\log2}{2}$ for any choice of boundary conditions $\a,\b$.

In the opposite regime, $R\to1$, the $\nu$ punctures in the right-hand side of \eqref{eq:RDMsigma_symmetric} collide tails first. Using the $\nu\nu$ OPE once more, only this time with bulk identity field $\id$, we immediately obtain the asymptotic behavior $S_n(\rho^{\sigma,S}_{\ab}) \stackrel{1-R\ll 1}{\approx} S_n(\rho^{\id}_{\ab})$ for any choice of boundary conditions. 
This concludes our argument, which extends \eqref{eq:numerics_sigma} to any replica index $n$. In fact, our reasoning is entirely independent of $n$, so that it persist in any analytic continuation in $n$.

%%%%%%%%%%%%%%%%%%%%%%%%%%%%%%%%%%
\textit{6. Discussion and Outlook}.--- In summary, we argued that in order to study entanglement in the presence of defects, the allowed space of states must be enlarged to include twisted sectors, \lref{M1}, and one must study quantum correlations of twisted excitations \lref{M2}, paying particular attention to the defect network. These modifications lead us to construct our main result, the RDMs \eqref{DefectRDMs}, with which we reproduce the numerically observed EEs in the Ising model for its non-trivial topological defects. Our construction manifestly avoids the problems \lref{P1} \& \lref{P2} of \eqref{eq:RDMold}. Moreover, our arguments are fairly general and apply to rational CFTs with non-diagonal spectrum, given appropriate adaptations in e.g. fusion products. 

We work in the boundary approach to entangling edges \cite{Ohmori_2015}, and extend it to the case where the states to be prepared require reference to a defect network. Indeed, the network is crucial, since it determines the allowed topological manipulations in replica manifolds. This is particularly important in the case of non-invertible defects, such as the duality defect above.

%The choice of boundary conditions influences the subleading contribution to entropy \cite{Ohmori_2015} by $\gf$-factors. In our examples, they play no particular role, since numerical results are present only for the EE \commentP{??}, and the $\gf$ can only be extracted upon comparing different R\'enyi entropies. 
Our results show that the choice of intermediate boundary condition $\a'$ in \eqref{DefectRDMs} is crucial. Indeed, our analysis rests on the circumstance that the defects can be unfused from the entangling edge and moved into the region $B$. This is possible for the \quotes{natural} choice $\a'=\a\fuse a$, which can be non-elementary. In the grouplike case, $a=\varepsilon$, this is the only choice. In the duality case $a=\sigma$, on the other hand, any $\a'\in \a\fuse\sigma$ may be chosen, and the defect cannot be unfused from the boundary a priori. One can furthermore investigate entanglement in junction fields, i.e. fields sitting at junctions of an arbitrary number of defects. Both options are left to future work. %\cn{should we be more specific and say \quotes{in an upcoming publication}?}.

It is furthermore interesting to consider RDMs for twisted states in other entanglement measures, such as relative entropy \cite{Lashkari_2014, Lashkari:2015dia}, entanglement negativity or reflected entropy, see \cite{LiuVertexStates} and references therein. Another candidate is symmetry-resolved entanglement \cite{Goldstein:2017bua} and asymmetry \cite{benini2025entanglementasymmetryhighernoninvertible}, which would make use of tube algebras \cite{Choi:2024tri}. As a final comment, we notice that our framework allows to consider entanglement through defects relating different phases of a CFT, for instance T-duality (order-disorder duality) in compact boson theories.

\bigskip

\textbf{Acknowledgements}.--- It is a pleasure to thank  Ananda Roy, Ingo Runkel, Martin Schnabl and Amartya Singh for fruitful discussions. CN's work is funded by the European Union’s Horizon
Europe Research and Innovation Programme under the Marie Sklodowska-Curie Actions
COFUND, Physics for Future, grant agreement No 101081515. PR is supported by European Structural and Investment Funds and the Czech Ministry of Education, Youth and Sports (Project FORTE CZ.02.01.01/00/22 008/0004632).
%\cn{Paolo, don't forget to thank your monetary overlords here ;P }

%\newpage
%\newpage

\bibliographystyle{apsrev4-2}
\bibliography{entanglement_entropy_biblio.bib}

\appendix

\begin{widetext}
\appendix
%%%%%%%%%%%%%%%%%%%%%%%%%%%%%%%%%%%%%%%%%%%%
\section{Twisted Annulus partition functions in the small $\epsilon$ limit}\label{appAnnulusZ}
%%%%%%%%%%%%%%%%%%%%%%%%%%%%%%%%%%%%%%%%%%%%
Consider an $n$-replica annulus containing a $2n$ twist fields $\phi$ spanning some defect network that does not connect to the boundary. In case of the trivial twist, these are just ordinary bulk fields. We furthermore allow for a single grouplike defect $D_g$ to stretch from one boundary of the annulus to the other. In the case that $\phi$ is an ordinary bulk field (trivial twist) and $g=\id$, this setup reduces to the standard replica geometry for excited states \cite{Ibáñez_Berganza_2012, Northe:2025qcv}. Further details about the topological defect network are not necessary for our purposes here -- only the conformal weights $h_\phi$ and $\bh_\phi$ matter.

Although they shall not be important for our purposes here, for completeness we mention the field insertion points \cite{Northe:2025qcv}. Fields stemming from the $m^\th$ RDM are inserted at 
\begin{equation}\label{wCoordnInsertions}
 w(\infty_m)
 =
 \iu\frac{\pi}{2\wi}(1-R+2(m-1))\,,
 \qquad
 w(0_m)
 =
 \iu\frac{\pi}{2\wi}(1+R+2(m-1))\,,
 \qquad
 m=1,\dots,n\,.
\end{equation} 
where $\infty_m$ is the image of $\infty$ and similarly for the origin and $\wi=2\log\left(\frac{2}{\epsilon}\sin(\pi R)\right)$.

Writing $\X_n^\phi(w)$ for the collection of twist fields, we thus consider the following trace
\begin{align}
\tr_\ab[g\,q^{nH_\ab}\X_n^\phi(w)]
=
(n\tau)^{-2nh_\phi}(n\btau)^{-2n\bh_\phi}
\bbra{\a_g}\tq^{\frac{H}{4n}}
\,\X_n^\phi(\tw)\,
\tq^{\frac{H}{4n}}\bket{\b_g}
\end{align}
where $q=e^{2\pi \iu \tau}$ is the modular nome with modular parameter $\tau=\iu\pi/\wi$. This equation is the standard duality between a boundary and bulk description (open-closed worldsheet duality) implemented by $w=n\tau\,\tw$ and $\tau\to\ttau=-1/\tau$. The transformed modular nome $\tq=e^{2\pi\iu\ttau}=e^{-2\wi}$ shrinks to zero in the limit $\epsilon\to0$, making it a good expansion variable. On the left-hand side propagation is driven by the BCFT Hamiltonian $H_\ab$, while on the right-hand side it is driven by the bulk Hamiltonian $H$. Note the unusual placement of the bulk propagators $\tq^{H/(4n)}$ emphasizing that the bulk fields are distributed uniformly in the center of the annulus. The prefactors on the right-hand side are the standard Jacobian associated with transforming the string of fields, $\X_n^\phi(w)\to\X_n^\phi(\tw)$. Finally, the $g$-twisted boundary states
\begin{equation}
    \bket{\a_g}
    =
    \sum_iB_{\a_g}^i\iket{i_g}\,,
    \qquad 
    (L_n-\bL_{-n})\iket{i_g}=0\,,
    \qquad
    \iket{i_g}\in\cH_i\otimes\cH_{i^+}
\end{equation}
are constructed from $g$-twisted Ishibashi states built on spinless primaries $i$ of the $g$-twisted bulk Hilbert space. We associate the conformal dimension $h_g(=\bh_g)$ to the lightest primary state $\ket{0_g}$ in this boundary state.  

In order to treat the $\epsilon\to0$ limit in the usual way, a resolution of the identity $\id_g=\sum_k\ketbra{k_g}{k_g}$, but now in the $g$-twisted sector, is appropriately inserted
\begin{align}
\bbra{\a_g}\tq^{\frac{H}{4n}}
\,\X_n^\phi(\tw)\,
\tq^{\frac{H}{4n}}\bket{\b_g}
&=
\sum_{k,l}\bbra{\a_g}\tq^{\frac{H}{4n}}
\ketbra{k_g}{k_g}
\,\X_n^\phi(\tw)\,
\ketbra{l_g}{l_g}
\tq^{\frac{H}{4n}}\bket{\b_g}\notag\\
&\overset{\tq\to0}{\simeq}
\tq^{\frac{h_g-\cc}{24n}}
B^{0}_{\a_g}B^{0}_{\a_g}\bra{0_g}\X_n^\phi(\tw)\ket{0_g}
\end{align}
%To be completely accurate here, we have inserted a CPT operator $\CPT$ on the left boundary, which accounts for the opposite orientations of the boundaries. This does not play a role in the main text however, since it mainly conjugates the representation label, and our representations are self-conjugate, $i=i^+$. 
In the limit $\epsilon\to0$ the boundaries are pushed away, so that the left-over correlator is that of a defect-dressed cylinder. In conclusion,
\begin{align}
Z_\ab^{\phi,g}(q^n)
:=
\tr_\ab[g\,q^{nH_\ab}\X_n^\phi(w)]
&\overset{\tq\to0}{\simeq}
(n\tau)^{-2nh_\phi}(n\btau)^{-2n\bh_\phi}
\tq^{\frac{h_g-\cc}{24n}}
B^{0}_{\a_g}B^{0}_{\b_g}\bra{0_g}\X_n^\phi(\tw)\ket{0_g}
\end{align}
For unitary theories we have $h_g\geq0$, with equality only for the lightest untwisted primary, which is the conventional vacuum field $\id$. In this case we revert to a \quotes{normal} cylinder and the g-factors are recovered, $B_{\a_\id}^0=\gf_\a$. The untwisted sector dominates sums of various such twisted annuli,
\begin{align}\label{appTwistedAnnulusPhiSum}
\sum_{g\in G}\Lambda_g
Z_\ab^{\phi,g}(q^n)
%\tr_\ab[g\,q^{nH_\ab}\X_n^\phi(w)]
%
&\overset{\tq\to0}{\simeq}
(n\tau)^{-2nh_\phi}(n\btau)^{-2n\bh_\phi}
\tq^{-\frac{\cc}{24n}}
\,
\Lambda_g\,\gf_\a\gf_\b\,
\underbrace{\bra{0}\X_n^\phi(\tw)\ket{0}}_{\corr{\X^\phi_n(\tw)}_{\mathrm{cyl}}},
\end{align}
for some constant coefficient $\Lambda_g$.

%%%%%%%%%%%%%%%%%%%%%%%%%%%%%%%%%%%%%%%%%%%%
\section{R\'enyi moments for disorder field $\mu$}
In the main text, we argue that symmetric RDM $\rho_\ab^{\mu,S}$ and interface RDM $\rho_\ab^{\mu,I}$ produce the same R\'enyi entropies, at least to leading order in $\epsilon$. Consider the following defect moves:
\begin{equation}\label{ReplicaDualityMove_app}
\vcenter{\hbox{\includegraphics[height=2.5cm]{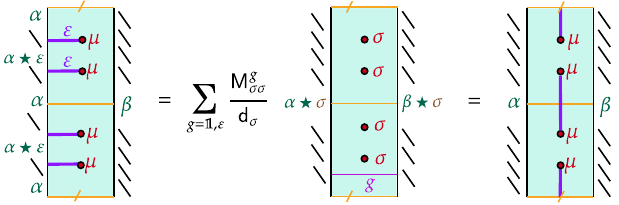}}}
%    \raisebox{-.45\height}{\includegraphics[scale=.2]{Pics/ReplicaDualityApp.png}}
\end{equation}
Writing $Z_\ab^{\mu,I}$ for the partition function on the left and $Z_\ab^{\mu,S}$ for the partition function on the right, these pictures are translated into equations by,
\begin{align}
Z_\ab^{\mu,I}(q^n)
=
Z_\ab^{\mu,S}(q^n)
&=
\sum_{g}\frac{\fusb_{\sigma\sigma}^g}{\qdim_\sigma}Z_{\a\fuse\sigma,\b\fuse\sigma}^{\sigma,g}(q^n)\notag\\
&\simeq
(n\tau)^{-2nh_\phi}(n\btau)^{-2n\bh_\phi}
\tq^{-\frac{\cc}{24n}}
\,\gf_\a\gf_\b\,
\corr{\X^\sigma_n(\tw)}_{\mathrm{cyl}}
\end{align}
which used \eqref{appTwistedAnnulusPhiSum}, $\fusb_{\sigma\sigma}^\id=\qdim_\sigma^{-1}$ and $\gf_{\a\fuse\sigma}=\qdim_\sigma\gf_\a$. Moments are now easily computed
\begin{align}
\tr_\ab[(\rho_\ab^{\mu,I})^n]
=
\tr_\ab[(\rho_\ab^{\mu,S})^n]
\simeq
\tr_\ab[(\rho_\ab^{\sigma})^n]
=
\tr_\ab[(\rho_\ab^{\id})^n]\cF_n^\sigma(R)\,,
\qquad
\cF^\sigma_n(R)
:=
\frac{1}{n^{4nh_\sigma}}\frac{\corr{\X^\sigma_n(\tw)}}{(\corr{\X_1^\sigma(\tw)})^n}
\end{align}
Because $\cF_n^\sigma(R)=1$ \cite{Ibáñez_Berganza_2012}, in the regime $\epsilon\to 0$ the entanglement entropies of $D_\varepsilon$-dressed RDMs equal that of the (untwisted) vacuum state $\rho_\ab^\id$,
\begin{equation}
    S_n(\rho_\ab^{\mu,S})
    =
    S_n(\rho_\ab^{\mu,I})
    \simeq
    S_n(\rho_\ab^{\id})
    =
    \frac{1+n}{n}
    \frac{\cc}{6}\log\left(\frac{\ell(R)}{\epsilon}\right)
    +
    \log(\gf_\a\gf_\b)
    +\dots
\end{equation}
as claimed in the main text.

\end{widetext}

\end{document}